\documentstyle[12pt,blois]{article}
\begin{document}

\heading{\bf  A WAY OUT OF THE DISK ANGULAR MOMENTUM CATASTROPHE IN
 HIERARCHICAL HYDRODYNAMICAL SIMULATIONS}

\author{ R. Dom\'{\i}nguez-Tenreiro$^1$, P.B. Tissera$^{1,2}$ and A. S\'aiz$^1$}
 {$^1$ Dpt. F\'{\i}sica Te\'orica, Univ. Aut\'onoma de Madrid,
   Cantoblanco, 28049 Madrid, Spain}
 {$^2$ IAFE, Casilla de Correos 67, Suc. 28, 1428 Buenos Aires, Argentina}

\begin{bloisabstract}
We present results that  suggest that DAMCs plaguing hierarchical
hydrodynamical simulations of galaxy formation, not including
star formation processes, are due to the inward gas transport that follows 
bar disk instabilities triggered by interactions and mergers.
They also show that DAMCs can be easily avoided by including star-forming
processes, as they lead unavoidably to the formation of compact stellar bulges
that stabilize disks against bars.
The formation of disks similar to those observed demands, in addition, that
not all the gas is depleted into stars at high $z$, so that they
can be formed  at lower $z$.
\end{bloisabstract}

\section{Disk Formation in Hierarchical  Hydrodynamical Simulations }

 According to Fall and Efstathiou's (FE) 
  standard model of disk formation \cite{FE80}, 
   extended disks similar to those
   observed in spiral galaxies can be formed from the
diffuse halo gas component provided that
{\it gas conserves its specific angular
    momentum (j) during collapse}.
However, so far, no hydrodynamical
simulation of galaxy formation in  fully consistent hierarchical
cosmological scenarios had been able
to produce extended disks
similar to observed spirals.
The problem was either
 the excessive loss of angular momentum by the gas clumps
as they merge inside the dark haloes,
when no star formation processes are considered,
resulting in too concentrated disks
  (i.e., the so-called {\it disk angular momentum catastrophe} 
  problem,
 hereafter DAMC  \cite{Nal95}
\cite{NS97} \cite{Wal98} and references quoted therein),
or the too early gas exhaustion into stars as it cools and
collapses, leaving no gas to form disks at low $z$ \cite{TDT98} 
\cite{SN98}.
In this paper we report on some results of disk formation in hierarchical
hydrodynamical simulations \cite{Tal97} \cite{TDT98} \cite{DTTS98}
where a simple implementation of 
star formation  that {\it prevents
 gas depletion at high redshifts}, 
 but {\it permits the formation of stellar bulges},
 has allowed extended and populated disks  to form  at later
 times.
We have followed the evolution of $64^3$ particles in a periodic box of
10 Mpc ($H_0 = $50 km s$^{-1}$ Mpc$^{-1}$) using a SPH code coupled to the
high resolution AP3M code  \cite{TC92}, either
{\it including a star formation algorithm} with star formation
efficiency $c = 0.01$ ({\bf S1} simulation)  or {\it not}
({\bf S2} simulation).
The initial distribution of positions and velocities is the same
in both S1 and S2, and is consistent with
a standard flat CDM cosmology, with $ \Omega_{\rm b} = 0.1,
\Lambda = 0 $ and $b = 2.5$. All, dark, gas and star particles have the same
mass, $M = 2.6 \times 10^8$ M$_{\odot}$.
The gravitational   softening length is 3 kpc and the minimum allowed
smoothing length is 1.5
kpc.
 Baryonic objects forming disk-like structures (DLOs) identified in S1
 have stellar bulge-like cores and extended, populated disks, their
 masses and specific angular momenta are compatible with those of observed
 spirals, and their bulge and disk scales, $R_{\rm b}$
 and $R_{\rm d}$, respectively, are also consistent with their
 observable values. 
 In S2, DLOs have an inner, rather disordered gas concentration and, also, 
 extended disks, but much less populated than their S1 counterparts.
 Gas particles inside their optical radii have too low $j$,
 and their  bulge and disk scales disagree with observations 
 (see Table 1 and \cite{F83}  \cite{C97} \cite{DTTS98}).

\medskip
\begin{center}
{\bf Table 1.} Some Characteristics of DLOs with $N_{\rm baryon} > 150$.
\end{center}
\begin{center}
\begin{tabular}{|l|l|l|l|l|l|l|l|l|l|l| }
\hline DLO & 1 & 2 & 3 & 4 & 5 & 6 & 7 & 8 & 9 & 10 \\ \hline 
 $N_{\rm gas}$ & 348& 359 &307 &311 &210 &151 &227 &189 & 108 & 109 \\
 $N_{\rm star}$& 278& 240 &211 &215 &95  &69  &79  &157 &99   &  47  \\
$R_{\rm b}$ (S1, kpc)&0.74&0.74&0.85&0.74&0.54&0.53&0.99&0.49&0.54&0.40\\
$R_{\rm b}$ (S2, kpc)&1.29&1.29&1.41&1.19& & & 1.27&1.32&1.47&1.23 \\
$R_{\rm d}$ (S1, kpc)&7.33&5.66&10.90&9.98&6.50&5.61&6.56&5.29&7.07&9.75\\
$R_{\rm d}$ (S2, kpc)&6.99&7.08&14.04&9.02& & & 5.87&5.62&13.31&10.75\\
\hline
\end{tabular}
\end{center}
\medskip

\begin{figure}[t]
\input epsf
\leavevmode
\epsfxsize=18.5cm

\epsfbox{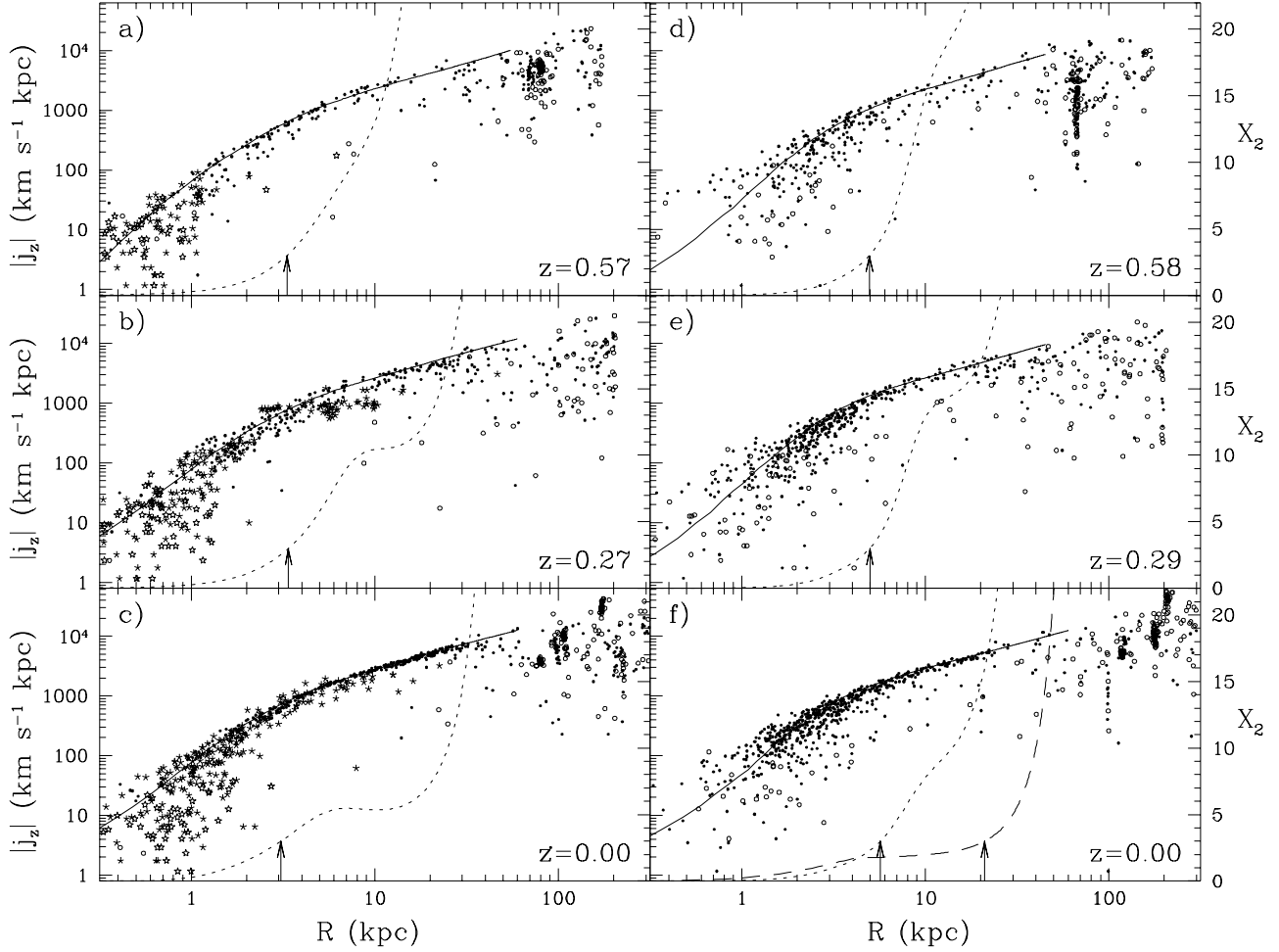}
\vspace*{-6.0cm}

\caption{
\baselineskip 2.8 true mm\footnotesize
Specific angular momentum component along $\vec{J}_{\rm dis}$ for each baryon
particle of halo \#1, versus their positions at different $z $.
{\it Points}: gas particles, {\it stars}: stellar particles;
{\it open symbols}: counterrotating particles.
{\it Left panels}: S1 version at different $z$; {\it right panels}: 
S2 version at
approximately the same $z$.
{\it Full lines}: $v_c(R) R$;
{\it dotted lines}: $X_2(R)$ for actual disks at each $z$;
{\it dashed line}: $X_2(R)$ for the pure exponential version at $z = 0$.
 {\it Arrows} mark  
$R_{\rm st}^{\rm ad}$ and $R_{\rm st}^{\rm ped}$,  where $X_2(R) = 3$.}
\end{figure}

So, stellar bulges seem to be critical to ensure global angular momentum 
conservation in the assembly of disks in hierarchical 
hydrodynamical cosmological simulations.
In fact, it is known that bulges play a fundamental role in stabilizing 
disks against the bar instability mode, that, otherwise, would cause a strong 
inward material  transport due to angular momentum losses \cite{CST95} 
\cite{MH96}
\cite{vdB98}.
To clarify their role, we briefly describe how disks are assembled
in S1  and S2 simulations
\cite{TDT98} \cite{DTTS98}.
i) First, dark matter haloes collapse at high $z$
forming a first generation of (very small) disks and stars. ii) Then, the first
unstabilizing mergers at high $z$ happen,
resulting in disk disruption and rapid
mass inflow to the central regions with angular momentum loss and violent star
formation, mainly at the central regions.
 Also, most preexisting stars will concentrate at the
 center of the new object through violent relaxation. These two processes help
 build up a central stellar bulge-like structure. iii) After the first mergers,
a
disk is regenerated through an infall of gas particles, either belonging to the
baryonic merging clumps or diffuse. For example, a
compact stellar bulge and an  almost cold disk in  DLO \#1 of S1 
at $z = 0.57$ are
apparent in Figure 1a.
iv) After disk regeneration, the system can undergo  new  major
merger events at lower $z$.
During the orbital decay phase, previous to the actual fusion of the DLOs, most
of their orbital angular momentum is transported to (the particle components of)
each host halo, spinning it up  (as in \cite{B92} \cite{BH}).
Because,
now, the disks involved in
the merger are stabilized by their bulges, no strong gas inflow occurs in this
phase (as in \cite{MH96}). As the disks approach one another,
they are heated and finally  disrupted, but the high
efficiency of gas shocking and cooling, and the symmetry of the central 
potential,
quickly puts those
of their gas particles
with high angular momentum into a new intermediate disk,
while their low angular
momentum particles sink to the center where most of them are transformed into
stars, feeding the bulge. The stellar bulge of the smaller DLO is eventually
destroyed and incomplete orbital angular momentum loss puts
most of its stars on the
remnant disk (Fig. 1b, note incomplete relaxation).
v) Relaxation and disk regeneration are completed. Most of disk external
particles are supplied by infall, as in iii) (Fig. 1c).

Note in Fig. 1c that at $z=0$ most gas particles placed at 
$R_i \stackrel{<}{\scriptstyle \sim}$ 30 kpc have
$j_{z,i} \simeq |\vec j_i| \simeq  v_{\rm c}(R_i)R_i$, where $v_{\rm c}(R) =
G M(<R)/R$,
with a small
dispersion around this value, that is, they follow circular
trajectories on the equatorial
plane, forming a cold thin disk. In contrast,
those at $R_i \stackrel{>}{\scriptstyle \sim} 30$ kpc (halo gas
particles) are disordered, with
their $|j_{z,i}|$ taking any value under the full line.
Roughly half of them are in counterrotation (i.e., with $j_{z,i} < 0$).
The specific angular momentum of halo gas particles, however, is the
same as that of gas disk particles, and, also, the same as that of 
dark matter particles.  
Stars at $R_i \stackrel{<}{\scriptstyle \sim} 2$ kpc form a compact central
relaxed core, with $\vec{j}_i$
without any preferred direction and very low $|\vec j_i|$
(that is, they have been formed from gas particles that had lost
much of their $|\vec j_i|$), while those at $R_i
\stackrel{>}{\scriptstyle \sim} 2$ kpc roughly follow a (thicker) disk.

The assembly of galactic-like objects in S2 follows the same stages. We recall
that in both simulations, haloes and merger trees are identical. The main
difference is that in S2, the i) and ii) stages do not result
in a stellar core,
and, consequently, in iii) stage an unstable gas disk is formed, susceptible
to grow bars. In particular, during the orbital decay phase in iv),
 strong gas inflow and $j$ loss are induced (i.e., a DAMC,
 see also Fig. 1d, and \cite{MH96} 
 \cite{BH}).
 The actual fusion completes the gas inflow (Fig. 1e),
  involving most of the gas particles originally in the merging disks.
  Few of them are left for disk regeneration, so that,  in phase v), new disks
  are formed almost only from halo gas particles (Fig. 1f), and hence 
  their low population.

  The behaviour patterns described so far are common to the other DLOs
   in S1 or S2.
  In any case,  particles in the external cold disk component 
  at $z = 0$ have fallen in the quiescent phase of evolution that
  follows the last major merger event,
  according with FE's scenario, 
  while most of those in the central regions (in S2), 
  or those giving rise to stars in the bulge (in S1) have been involved in 
  a DAMC. Most particles in the intermediate disk component in S1 DLOs
  belonged to the merging objects  and have suffered a partial angular
  momentum conservation in the merger event.
  Hence, cold thin disks naturally appear in the non-violent 
  phases of evolution.
  However, as stated, cold disks are strongly unstable against the bar mode. 
  Some works on disk stability (\cite{CST95} \cite{vdB98} and
  references therein) suggest that
  sometimes a central bulge is needed to ensure stability, as massive
  dark haloes are not always able to stabilize a given amount of 
  baryons as pure exponential disks ({\it ped}s). 
  This could be the process at work
  in  DAMCs observed in hydrodynamical simulations.
  To find out  whether this is the case here, we have calculated the 
  $X_2(R)$ parameter \cite{T81} \cite{BT87} for the disk component of
  our DLOs at different $z$ (Fig. 1), and, also, for their 
  {\it ped} versions (i.e., putting all their respective baryonic masses
  distributed as a {\it ped}).
  Recalling the $X_2(R)$ stability criterion, if we define the 
  stability thresholds, $R_{\rm st}^{\rm ad}$ and $R_{\rm st}^{\rm ped}$,
  as the points where $X_2(R)=3$ for actual and pure exponential disks,
  respectively, it is apparent from  Fig. 1 that disks, when present, are
  stable: they are
  detected at $R > R_{\rm st}^{\rm ad}$
  if they have had enough time  to form after the last merger.
 By contrast, the {\it ped} version of DLO \#1
 at $z = 0$ would be stable only at larger $R$ ($R >
 R_{\rm st}^{\rm ped} \simeq 21$ kpc).
This behaviour is
common to any DLO in S1 or S2, and so central mass concentrations are needed
to stabilize these disks. 
These results strongly suggest that DAMCs result from strong gas inflows 
due to disk instabilities triggered by interactions and mergers during
the assembly of galaxy-like objects, and that they can be easily avoided 
by stabilizing  the disks with stellar bulges.

\acknowledgements{We are indebted to the DGES (Spain) for financial support.
P.B. Tissera thanks the Astrophysics Group at ICSTM (London)
for their hospitality.}

\vfill
\end{document}